\documentclass[aps,prb,twocolumn,superscriptaddress,floatfix,longbibliography]{revtex4-2}

\usepackage{amsmath}
\usepackage{graphicx}
\usepackage{dcolumn}
\usepackage{bm}
\usepackage{bbold}
\usepackage{hyperref}
\usepackage{xcolor}

\begin{document}

\preprint{APS/123-QED}

\title{Quantum dynamic response-based NV-diamond magnetometry: Robustness to decoherence and applications in motion detection of magnetic nanoparticles}

\author{Wenkui Ding}
\affiliation{Department of Physics, Zhejiang Sci-Tech University, 310018 Zhejiang, China}
\author{Xingyu Zhang}
\affiliation{Department of Physics,
Xiamen University, 361005 Fujian, China}
\author{Jing Liu}
\affiliation{MOE Key Laboratory of Fundamental Physical Quantities Measurement, National Precise Gravity Measurement Facility, School of Physics, Huazhong University of Science and Technology, Wuhan 430074, China}
\author{Xiaoguang Wang}
\email{xgwang@zstu.edu.cn}
\affiliation{Department of Physics, Zhejiang Sci-Tech University, 310018 Zhejiang, China}

\date{\today}

\begin{abstract}
We propose a novel quantum sensing protocol that leverages the dynamical response of physical observables to quenches in quantum systems.
Specifically, we use the nitrogen-vacancy (NV) color center in diamond to realize both scalar and vector magnetometry via quantum response. Furthermore, we suggest a method for detecting the motion of magnetic nanoparticles, which is challenging with conventional interference-based sensors. To achieve this, we derive the closed exact form of the Berry curvature corresponding to NV centers and design quenching protocols to extract the Berry curvature via dynamical response.
By constructing and solving non-linear equations, the magnetic field and instantaneous motion velocity of the magnetic nanoparticle can be deduced.
We investigate the feasibility of our sensing scheme in the presence of decoherence and show through numerical simulations that it is robust to decoherence.
Intriguingly, we have observed that a vanishing nuclear spin polarization in diamond actually benefits our dynamic sensing scheme, which stands in contrast to conventional Ramsey-based schemes.
In comparison to Ramsey-based sensing schemes, our proposed scheme can sense an arbitrary time-dependent magnetic field, as long as its time dependence is nearly adiabatic.
\end{abstract}

\maketitle


\section{Introduction}
Quantum metrology~\cite{braunstein1994statistical,giovannetti2006quantum,giovannetti2011advances,escher2011general,pezze2018quantum,braun2018quantum} and quantum sensing~\cite{degen2017quantum,barry2020sensitivity} have attracted significant attention in recent years.
Quantum sensors, leveraging the unique properties of quantum systems, hold promise for detecting weak or nanoscale signals that surpass the capabilities of classical sensors. 
While most quantum sensors rely on interference schemes, there are situations where implementing interferometry or Ramsey-based schemes becomes challenging~\cite{yurke1986su,wang2022sensing}. 
One such scenario arises when the signal to be detected exhibits a short period of viability, making it impractical to accumulate the necessary phase for information encoding in the interference-based scheme~\cite{maclaurin2013nanoscale}. As a result, there is a growing emphasis on exploring novel mechanisms to realize innovative quantum sensing schemes, driving rapid developments in the field of quantum science and technology~\cite{braun2018quantum,budich2020non,review2020jan,chu2020quantum,mishra2021driving,ding2022dynamic}.

In recent studies~\cite{Gritsev6457,rigolin2008beyond,avron2011quantum,de2010adiabatic,de2013microscopic}, the concept of dynamical response has been proposed as a means to detect geometric quantities in quantum many-body systems. Notably, the emergence of Berry curvature in the nonadiabatic response of physical observables to slow quenches, irrespective of the system's interaction nature, has been identified~\cite{Gritsev6457}. Building upon these findings, our study showcases the potential of utilizing the mechanism of dynamic response for quantum sensing, offering a complementary approach to the conventional interference-based sensing schemes. Specifically, we present quantum response-based sensing schemes utilizing nitrogen-vacancy (NV) color centers in diamond~\cite{doherty2012theory,doherty2013nitrogen}.
The NV center in diamond is a highly attractive candidate for quantum sensing due to its efficient initialization and readout capabilities through optical excitations, as well as its relatively long coherence time, even at ambient temperature~\cite{balasubramanian2009ultralong,hanson2008coherent}.
Consequently, extensive theoretical and experimental investigations have been conducted to explore the quantum sensing potential of NV centers~\cite{wood2018limited,liu2019nanoscale,ajoy2015atomic,mamin2013nanoscale,grinolds2013nanoscale,acosta2009diamonds}.
Notably, NV centers have demonstrated the ability to sense magnetic fields with nanoscale spatial resolution~\cite{maze2008nanoscale,grinolds2013nanoscale}.
Besides, the geometric quantity, like the geometric phase, in NV centers has been investigated~\cite{maclaurin2012measurable,yu2020experimental} and proposed in the applications in quantum sensing, like gyroscope~\cite{ledbetter2012gyroscopes,ajoy2012stable} and magnetometer~\cite{arai2018geometric}.
Furthermore, owing to diamond's chemical inertness and the excellent quantum property under ambient condition, NV sensors hold promise for applications in bioimaging~\cite{balasubramanian2014nitrogen}.
In this study, we propose a novel approach using NV centers through quantum dynamic response to sense the motion of magnetic nanoparticles, which has the potential to find applications in the field of bioimaging.

Before introducing the dynamic response-based sensing scheme, we provide a brief overview of the quantum response theory~\cite{Gritsev6457,de2010adiabatic}.
By employing adiabatic perturbation theory~\cite{rigolin2008beyond}, the general formula for quantum response can be derived as follows (see Appendix A for detailed information):
\begin{equation}
\label{eq:qr_formula_0}
M_\mu=\text{const}+v_\lambda\mathcal{F}^{(m)}_{\mu\lambda}+\mathcal{O}(v_\lambda^2).
\end{equation}
Here, $M_\mu$ represents the observable being measured in the experiment, often referred to as the generalized force along the $\mu$-direction. It can be defined as $M_\mu\equiv-\langle \psi(t_f)|\partial_\mu H|\psi(t_f)\rangle$, with $\partial_\mu H\equiv\partial H/\partial \mu$. 
The quantum state evolves according to $|\psi(t_f)\rangle=\mathcal{T}e^{-i\int_0^{t_f}H(t^\prime)dt^\prime}|\Psi_m(0)\rangle$, where $\mathcal{T}$ denotes the time-ordering operator and the time dependence of the Hamiltonian is introduced by the time-dependent parameters, $H(t)=H(\lambda(t),\mu(t),\dots)\equiv H(\lambda,\mu,\dots)$.
The initial state is prepared as one of the instantaneous eigenstates of the Hamiltonian, $H(t)|\Psi_m(t)\rangle=E_m(t)|\Psi_m(t)\rangle$.
The quench process is achieved by varying the parameter $\lambda(t)$ over time, with $v_\lambda\equiv {\partial \lambda}/{\partial t}$ representing the instantaneous quench velocity along the $\lambda$-direction at time $t_f$.
Notably, the Berry curvature $\mathcal{F}^{(m)}_{\mu\lambda}$ corresponding to the instantaneous eigenstate $|\Psi_m(t_f)\rangle$, emerges as the coefficient in the non-adiabatic response when the quench velocity approaches zero.

We would like to make some comments regarding the sensor utility of the quantum response formula presented in Eq.~(\ref{eq:qr_formula_0}).
The validity of this equation does not rely on the specific details of the quench process, as long as the quench is performed in a nearly adiabatic manner.
Most notably, this formula indicates that by implementing quenches along the $\lambda$-direction and measuring the corresponding response along the $\mu$-direction, we can extract the value of the Berry curvature $\mathcal{F}_{\mu\lambda}^{(m)}$.
Since the Berry curvature is a geometric quantity solely determined by the parameter-dependent instantaneous eigenstate of the quantum system, it remains independent of the specific details of the quench process.
Moreover, if the physical quantity of interest is encoded within the Berry curvature, we can determine its value by measuring the Berry curvature using the quench-response mechanism.
Conversely, if the Berry curvature is known a priori, we can determine the instantaneous quench velocity by measuring the system's response.

This article is organized as follows:
In Sec. II, we derive the closed exact form of the Berry curvature associated with NV centers.
In Sec. III, we present a concrete dynamic response-based scheme for NV-magnetometry and assess its feasibility by considering the effects of decoherence.
In Sec. IV, we propose a protocol for detecting the motion of a magnetic nanoparticle using quantum response.
In Sec. V, we discuss the sensitivity of our dynamic sensing scheme.
Finally, summaries are made in Sec. VI.

\section{Berry curvature of NV centers}
Our focus is on utilizing the NV center in diamond to realize dynamic response-based quantum sensing. In this section, we aim to derive the analytical expression for the Berry curvature associated with this quantum system.
Generally, the Berry curvature can be expressed as the imaginary part of the geometric tensor, $\mathcal{F}_{\mu\lambda}=-2\Im[\chi_{\mu\lambda}]$.
The geometric tensor $\chi_{\mu\lambda}$ is defined as follows~\cite{provost1980riemannian}:
\begin{equation}
\label{eq:geometric_tensor_formula}
\chi_{\mu\lambda}=\langle\partial_\mu\Psi|\partial_\lambda\Psi\rangle-\langle\partial_\mu\Psi|\Psi\rangle\langle\Psi|\partial_\lambda\Psi\rangle.
\end{equation}
where $|\partial_\lambda\Psi\rangle\equiv \frac{\partial|\Psi\rangle}{\partial\lambda}$, and $|\Psi\rangle\equiv|\Psi(\lambda,\mu)\rangle$ represents a parameter-dependent quantum state.
In particular, when the parameter-dependent quantum state corresponds to the instantaneous eigenstates of the parameter-dependent Hamiltonian, given by $H(\lambda,\mu)|\phi_m(\lambda,\mu)\rangle=E_m(\lambda,\mu)|\phi_m(\lambda,\mu)\rangle$, the Berry curvature can be determined using the following expression:
\begin{equation}
\label{eq:berry_curvature_general}
\mathcal{F}_{\mu\lambda}^{(m)}=i\sum_{n\neq m}\frac{\langle \phi_m|{\partial_\mu H}|\phi_n\rangle\langle \phi_n|\partial_\lambda H|\phi_m\rangle-\mu\leftrightarrow\lambda}{\left[E_n(\lambda,\mu)-E_m(\lambda,\mu)\right]^2},
\end{equation}
assuming the eigenstate $|\phi_m\rangle$ is non-degenerate.

The Hamiltonian that describes the NV center driven by a time-varying magnetic field is given by~\cite{chen2018quantummetrology}:
\begin{equation}
\label{eq:main_Hamiltonian}
H(t)=DS_z^2+E(S_x^2-S_y^2)+g_e\mu_B\mathbf{h}(t)\cdot\mathbf{S}+\mathbf{S}\cdot\sum_{k=1}^N\mathbf{A}_k\cdot\mathbf{I}_k.
\end{equation}
Here, $\mathbf{S}=(S_x,S_y,S_z)$ represents the spin operator of the NV electronic spin, which has a spin quantum number $S=1$.
The Hamiltonian contains several important terms:
The first term represents the diagonal term of the zero-field splitting and $D\approx 2.87$ GHz, represents the zero-field splitting parameter, which exhibits temperature dependence and can be exploited for temperature sensing.
The second term corresponds to the off-diagonal term of the zero-field splitting, which captures the interaction between the NV center's electronic spin and an external electric field or stress, providing a means for electric field and stress detection~\cite{dolde2011electric}.
The third term corresponds to the Zeeman energy of the NV electronic spin in the presence of a time-varying magnetic field $\mathbf{h}(t)=(h_x(t),h_y(t),h_z(t))$, while $g_e$ is the NV electronic $g$-factor and $\mu_B$ is the Bohr magneton. This term enables the sensing of magnetic fields.
The last term describes the hyperfine interaction between the NV electronic spin and the surrounding nuclear spins, such as $^{13}$C nuclear spins with a spin quantum number $I=1/2$.
This term enables spin-sensing, where $\mathbf{I}_k$ represents the spin operator of the $k$-th nucleus and $\mathbf{A}_k$ represents the coupling strength of the NV electronic spin and the $k$-th nuclear spin.
The NV center in diamond possesses remarkable quantum properties, making it a versatile and promising quantum sensor under ambient temperature. 
While the last term is typically considered as the origin of decoherence of the NV electronic spin, for the purpose of demonstrating our dynamic response-based sensing protocol, we temporarily neglect this coupling term. Its effect will be carefully investigated in the subsequent section.

By neglecting the last coupling term in Eq.~(\ref{eq:main_Hamiltonian}), the simplified Hamiltonian can be expressed (by assuming $g_e\mu_B=1$ for simplicity) as follows: 
\begin{equation}
\label{eq:main_hamiltonian_matrix}
H=\begin{pmatrix}
D+h_z & \frac{h_x-ih_y}{\sqrt{2}} & E \\
\frac{h_x+ih_y}{\sqrt{2}} & 0 & \frac{h_x-ih_y}{\sqrt{2}} \\
E & \frac{h_x+ih_y}{\sqrt{2}} & D-h_z
\end{pmatrix}.
\end{equation}
To obtain the analytic form of the Berry curvature using Eq.~(\ref{eq:berry_curvature_general}), we need to find the explicit eigenstates and eigenvalues of this parameter-dependent Hamiltonian.
Fortunately, for a general $3\times 3$ Hermitian matrix~\cite{siddique2020eigenvalues,smith1961eigenvalues}, since all the eigenvalues are real, we can analytically calculate them in terms of the trignometric solutions (see Appendix B for more details).
To be specific, the eigenvalues of the Hamiltonian in Eq.~(\ref{eq:main_hamiltonian_matrix}) can be obtained as follows:
\begin{equation}
\label{eq:general_eigenvalue}
\begin{aligned}
E_1&=\frac{2}{3}\left[D-\Delta_0\cos{(\frac{\varphi-\pi}{3})}\right],\\
E_2&=\frac{2}{3}\left[D-\Delta_0\cos{(\frac{\varphi+\pi}{3})}\right],\\
E_3&=\frac{2}{3}\left[D+\Delta_0\cos{(\frac{\varphi}{3})}\right],\\
\end{aligned}
\end{equation}
where $\Delta_0\equiv\sqrt{\frac{3}{2}\text{Tr}[\mathcal{H}^2]}$ and $\cos\varphi=\frac{1}{2}\left(\frac{3}{\Delta_0}\right)^3\det(\mathcal{H})$.
Here, we have used the traceless Hamiltonian $\mathcal{H}\equiv H-\frac{\text{Tr}[H]}{3}\mathbf{1}$, with
\begin{equation*}
\begin{aligned}
\text{Tr}[\mathcal{H}^2]=&\frac{2}{3}D^2+2E^2+2h^2,\\
\det{(\mathcal{H})}=&\frac{2D}{3}(E^2+h^2-\frac{D^2}{9})+h_x^2(E-D)-h_y^2(E+D),
\end{aligned}
\end{equation*}
where $h^2=h_x^2+h_y^2+h_z^2$.
An obvious advantage of this trigonometric analytic form is that, it immediately reveals $E_1\leq E_2\leq E_3$ since $0\leq\varphi\leq \pi$.
To the best of our knowledge, the exact form of the eigenenergy of the NV Hamiltonian presented in this study has not been utilized in the existing literature. Conventionally, discussions on the eigenenergies or transitions of the NV center, a 3-level system, often rely on perturbation methods to obtain approximate results~\cite{barry2020sensitivity,doherty2012theory}. However, these approximate approaches can pose challenges when it comes to calculating the Berry curvature, which requires a more precise understanding of the system's eigenenergy structure.

The instantaneous eigenstates of the $3\times 3$ Hermitian matrix can be represented as the cross product of two three-dimensional vectors~\cite{kopp2008efficient}, $|\tilde{\Psi}_m\rangle=[(\mathbf{h}_1-E_m\mathbf{e}_1)\times(\mathbf{h}_3-E_m\mathbf{e}_3)]^*$, where $\mathbf{h}_j$ is the $j-$th column of the Hamiltonian and $\mathbf{e}_i$ is the unit vector, like $\mathbf{e}_1=(1,0,0)^{\text{T}}$.
Consequently, the explicit form of the instantaneous eigenstate corresponding to $E_m$ is given by
\begin{equation}
|\Psi_m\rangle=\frac{1}{\sqrt{2\mathcal{N}_m}}\begin{pmatrix}
{-E(h_x+ih_y)+(D_m-h_z)(h_x-ih_y)}\\
\sqrt{2}(-D_m^2+E^2+h_z^2)\\
{-E(h_x-ih_y)+(D_m+h_z)(h_x+ih_y)}
\end{pmatrix},
\end{equation}
where we have defined $D_m\equiv D-E_m$ and
\begin{equation}
\begin{aligned}
\mathcal{N}_m=D_m^4+&(h^2-2E^2-3h_z^2)D_m^2+2E(h_y^2-h_x^2)D_m\\
&+(E^2+h_z^2)(E^2+h^2),
\end{aligned}
\end{equation}
is the corresponding normalization factor.

Equipped with these exact eigenenergies and eigenstates, we can now calculate the Berry curvature corresponding to the eigenstate $|\Psi_m\rangle$ using Eq.~(\ref{eq:berry_curvature_general}).
While the Berry curvature is fundamentally determined by Eq.~(\ref{eq:geometric_tensor_formula}) once the explicit form of the parameter-dependent eigenstate is known, however, directly applying this equation poses challenges due to the intricacy of computing wave function derivatives.
Hence, we resort to the alternative expression provided by Eq.~(\ref{eq:berry_curvature_general}), which enables us to determine the Berry curvature without explicitly calculating the wave function derivatives.
After performing the involved yet straightforward calculations, we obtain the analytical expression for the Berry curvature associated with NV centers when the Cartesian components of the magnetic field are utilized as the driven parameters. The explicit forms of the Berry curvature components are given as follows:
\begin{widetext}
\begin{equation}
\label{eq:berry_curvature_cartesian}
\begin{aligned}
\mathcal{F}_{xy}^{(m)}&=\sum_{n\neq m}\frac{-2h_z(D_m+D_n)}{\mathcal{N}_m\mathcal{N}_n(D_m-D_n)}\Big[(D_m^++D_n^+)(D_m^-D_n^-h_x^2-h_y^2h_z^2)+(D_m^-+D_n^-)(D_m^+D_n^+h_y^2-h_x^2h_z^2)\Big],\\
\mathcal{F}_{xz}^{(m)}&=\sum_{n\neq m}\frac{2h_y(D_m^++D_n^+)}{\mathcal{N}_m\mathcal{N}_n(D_m-D_n)}\Big[2E(D_m^-D_n^-+h_z^2)h_x^2-(D_m+D_n)(h^2-h_z^2)h_z^2\Big],\\
\mathcal{F}_{yz}^{(m)}&=\sum_{n\neq m}\frac{2h_x(D_m^-+D_n^-)}{\mathcal{N}_m\mathcal{N}_n(D_m-D_n)}\Big[2E(D_m^+D_n^++h_z^2)h_y^2+(D_m+D_n)(h^2-h_z^2)h_z^2\Big],\\
\end{aligned}
\end{equation}
\end{widetext}
where we have introduced the notation $D_m^{\pm}\equiv D_m\pm E$.
It is worth noting that these analytical results reveal some intriguing features. Specifically, when $h_z=0$, we have $\mathcal{F}_{xy}=0$, and similarly, when $E=0$ and $h_z=0$, we find $\mathcal{F}_{xz}=\mathcal{F}_{yz}=0$.

With the explicit formulation of the Berry curvature at our disposal, we are now equipped to develop sensing protocols that harness the quantum dynamical response mechanism described by Eq.~(\ref{eq:qr_formula_0}). 
In the subsequent sections, we will illustrate specific sensing schemes based on quantum response and thoroughly examine their feasibility. Through these investigations, we aim to establish the practicality and effectiveness of employing the quantum dynamical response for sensing applications.

\section{scalar magnetometry and the robustness to decoherence\label{sec:scalar_rotating}}
\subsection{dynamic response-based sensing scheme using the rotating quench field\label{sec:basic_scheme}}
In this subsection, we present a specific scheme for scalar magnetometry utilizing the quantum response, focusing on a \textit{rotating} quench protocol. Furthermore, we consider the simplified scenario where $E=0$, which allows for a clear and concise presentation of the sensing procedure. 
Under these conditions, the Hamiltonian governing the dynamics of the NV center, driven by a magnetic field $\mathbf{h}(t)=h(\sin\theta\cos\phi,\sin\theta\sin\phi,\cos\theta)$, takes the form:
\begin{equation}
\label{eq:simple_Hamiltonian}
\begin{aligned}
H=DS_z^2+e^{-i\phi S_z}e^{-i\theta S_y}S_ze^{i\theta S_y}e^{i\phi S_z},
\end{aligned}
\end{equation}
where we have adopted the convention of rescaling the zero-field coupling strength by setting $h=1$, effectively incorporating it into the parameter $D/h\rightarrow D$.
Utilizing the eigenenergy expression derived in Eq.~(\ref{eq:general_eigenvalue}), we can explicitly calculate the eigenenergies as follows:
\begin{equation}
\label{eq:eigenvalue_s}
\begin{aligned}
E_1&=\frac{2}{3}\left[D-\sqrt{D^2+3}\cos(\frac{\varphi-\pi}{3})\right],\\
E_2&=\frac{2}{3}\left[D-\sqrt{D^2+3}\cos(\frac{\varphi+\pi}{3})\right],\\
E_3&=\frac{2}{3}\left[D+\sqrt{D^2+3}\cos(\frac{\varphi}{3})\right],\\
\end{aligned}
\end{equation}
where
\begin{equation}
\cos\varphi=\frac{D(-9-2D^2+27\cos^2{\theta})}{2\sqrt{(D^2+3)^3}}.
\end{equation}
Notably, due to the commutation relation $[e^{-i\phi S_z},H]=0$, the eigenenergies do not depend on the value of $\phi$.
Furthermore, the corresponding eigenstates can be obtained as follows:
\begin{equation}
\label{eq:eigenstate_s}
|\Psi_m\rangle=\frac{1}{\sqrt{\mathcal{N}_m}}\left(
\begin{array}{ccc}
e^{-i\phi}\sin\theta(D_m-\cos\theta)\\
\sqrt{2}(\cos^2\theta-D_m^2)\\
e^{i\phi}\sin\theta(D_m+\cos\theta)
\end{array}\right),
\end{equation}
where $D_m\equiv D-E_m$ and the normalization factor,
\begin{equation}
\mathcal{N}_m=2D_m^4-D_m^2+(1-3D_m^2)\cos{2\theta}+1.
\end{equation}
It should be noted that the analytic form of the eigenstate given in Eq.~(\ref{eq:eigenstate_s}) is not applicable when $\theta=\pi/2$ and $E_m=D$ (see Appendix B for more details). In fact, when $\theta=\pi/2$, the exact eigenvalues can be further simplified as $E_1=(D-\sqrt{4+D^2})/2$, $E_2=D$, and $E_3=(D+\sqrt{4+D^2})/2$, while the eigenstate corresponding to $E_2$ is represented by $|\Psi_2\rangle=1/\sqrt{2}(e^{-i\phi},0,-e^{i\phi})^{\text{T}}$.

Having obtained the explicit form of the eigenenergies and eigenstates, we can now proceed to calculate the Berry curvature using the formula in Eq.~(\ref{eq:berry_curvature_general}), where the derivatives of the Hamiltonian with respect to $\phi$ and $\theta$ are given by
\begin{equation}
\begin{aligned}
\partial_\phi H&=-\sin\theta \sin\phi S_x+\sin\theta \cos\phi S_y,\\
\partial_\theta H&=\cos\theta \cos\phi S_x+\cos\theta \sin\phi S_y-\sin\theta S_z.
\end{aligned}
\end{equation}
Utilizing the analytic form of the eigenstates presented in Eq.~(\ref{eq:eigenstate_s}), we obtain
\begin{equation}
\begin{aligned}
\langle \Psi_m|\partial_\phi H|\Psi_n\rangle&=\frac{2i}{\sqrt{\mathcal{N}_m\mathcal{N}_n}}(D_m^2-D_n^2)\sin^2\theta\cos\theta,\\
\langle \Psi_n|\partial_\theta H|\Psi_m\rangle&=\frac{1}{\sqrt{\mathcal{N}_m\mathcal{N}_n}}(D_n+D_m)(1-D_nD_m)\sin{2\theta}.
\end{aligned}
\end{equation}
Applying Eq.~(\ref{eq:berry_curvature_general}), we immediately observe that $\mathcal{F}_{\phi\phi}=\mathcal{F}_{\theta\theta}=0$, while $\mathcal{F}_{\phi\theta}^{(m)}=-\mathcal{F}_{\theta\phi}^{(m)}$.
In particular, the explicit form of the Berry curvature corresponding to the eigenstate $|\Psi_m\rangle$ is given by
\begin{equation}
\label{eq:simple_BC}
\begin{aligned}
\mathcal{F}_{\phi\theta}^{(m)}&=8\sin^3\theta\cos^2\theta\sum_{n\neq m}\frac{(D_n+D_m)^2(1-D_nD_m)}{\mathcal{N}_m\mathcal{N}_n(D_n-D_m)}.
\end{aligned}
\end{equation}
It is worth noting that $\mathcal{F}_{\phi\theta}^{(m)}$ is also independent of $\phi$.
Furthermore, we can express the Berry curvature for the ground state as follows:
\begin{equation}
\label{eq:berry_curvature_spherical}
\begin{aligned}
\mathcal{F}_{\phi\theta}^{(1)}&=8\sin^3\theta\cos^2\theta\times\\
&\left[\frac{(D_1+D_2)^2(1-D_1D_2)}{\mathcal{N}_1\mathcal{N}_2(D_2-D_1)}+\frac{(D_1+D_3)^2(1-D_1D_3)}{\mathcal{N}_1\mathcal{N}_3(D_3-D_1)}\right].
\end{aligned}
\end{equation}
Moreover, in the special case when $\theta=\pi/2$, the Berry curvature assumes a more compact form:
\begin{equation}
\mathcal{F}^{(1)}_{\phi\theta}(\phi,\theta=\frac{\pi}{2})=D-\frac{D^2+2}{\sqrt{D^2+4}}.
\end{equation}

\begin{figure}
\includegraphics[width=0.5\textwidth]{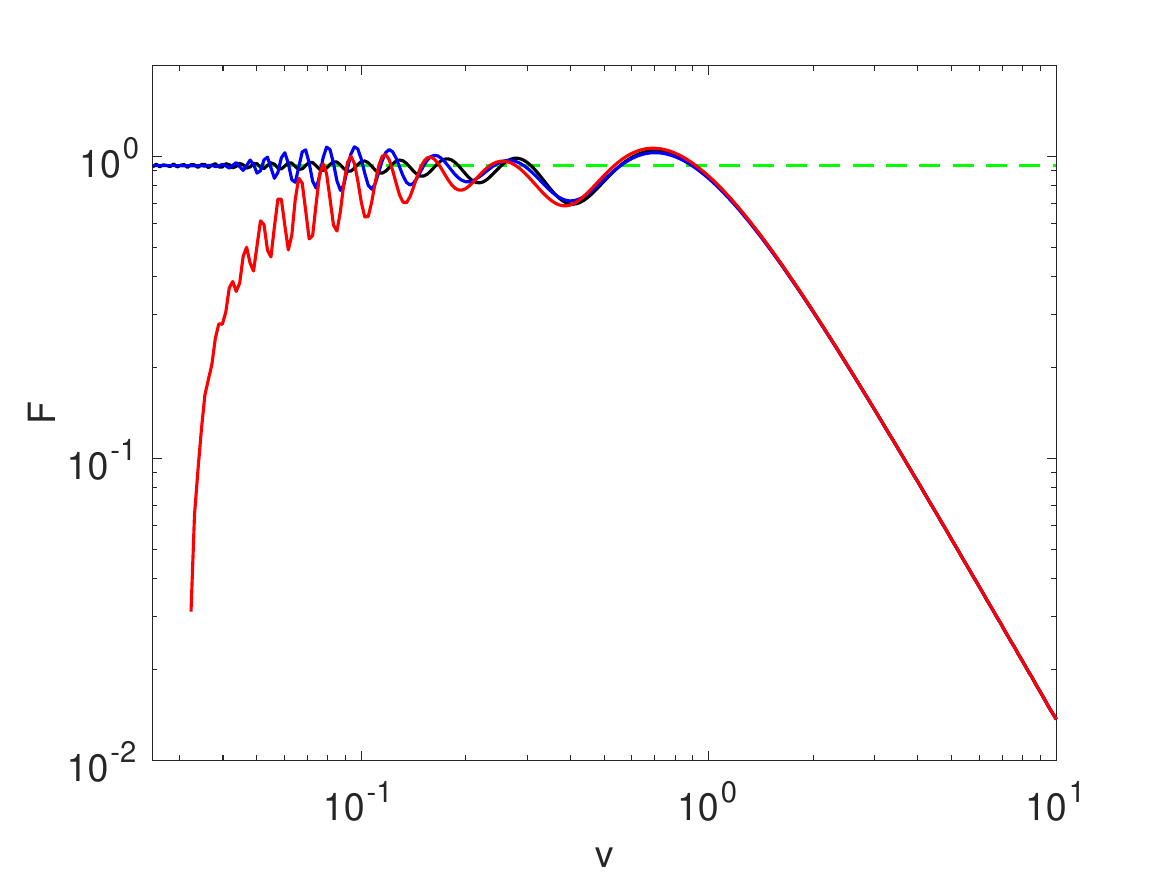}
\centering
\caption{\label{fig:qr}(Color online) Retrieval of the Berry curvature via quantum dynamical response in the presence of decoherence.
The quench is realized via the rotating magnetic field in Eq.~(\ref{eq:rotating_quench_field}), where $v$ is the quench velocity.
The dynamics of the system is governed by the Hamiltonian in Eq.~(\ref{eq:main_Hamiltonian}), where $N=20$ nuclear spins are considered, with the coupling strength $A_k=A=0.02$.
The parameters $D=0.06765$ and $E=0$ are also considered.
The retrieved Berry curvature is denoted by $F\equiv \langle S_y\rangle/v$.
We compare the results for two scenarios: one with zero nuclear polarization ($P=0$) represented by the blue solid line, and the other with a nuclear polarization of $P=0.2$ depicted by the red solid line. For reference, the green dashed line corresponds to the analytic result of the Berry curvature, while the black solid line represents the retrieved Berry curvature without considering the effects of decoherence.}
\end{figure}

We now present a concrete quenching protocol to demonstrate how the quantum response-based sensing scheme operates. We apply a rotating quench field given by
\begin{equation}
\label{eq:rotating_quench_field}
\begin{array}{ccc}
h_x(t)=\sin\left(\frac{v^2t^2}{2\pi}\right),& h_y(t)=0, & h_z(t)=\cos\left(\frac{v^2t^2}{2\pi}\right),
\end{array}
\end{equation}
where the quench is realized through $\theta(t)=\frac{v^2t^2}{2\pi}$.
This choice of the rotating quench ensures that the driving at the initial time is adiabatic since $v_\theta(t=0)=0$.
Specifically, we measure the response $\langle \partial_\phi H\rangle$ at $t_f=\pi/v$ with an instantaneous quench velocity of $v_\theta(t_f)=v$.
Firstly, we perform numerical simulations to verify the validity of the quantum response formula stated in Eq.~(\ref{eq:qr_formula_0}), which asserts that
\begin{equation}
\langle \psi(t_f)|\partial_\phi H|\psi(t_f)\rangle\approx \langle \Psi_1(0)|\partial_\phi H|\Psi_1(0)\rangle+ v_\theta\mathcal{F}_{\phi\theta}^{(1)},
\end{equation}
where $|\psi(t_f)\rangle=\mathcal{T}e^{-i\int_0^{t_f}H(t^\prime)dt^\prime}|\Psi_1(0)\rangle$.
Specifically, for the rotating quench protocol given by Eq.~(\ref{eq:rotating_quench_field}), we aim to verify that
\begin{equation}
\frac{\langle \psi(t_f)|S_y|\psi(t_f)\rangle}{v} \approx \frac{D^2+2}{\sqrt{D^2+4}}-D.
\end{equation}
To accomplish this, we solve the time-dependent Schrödinger equation to obtain the left-hand side of the equation. The result is depicted as the black solid line in Fig.~\ref{fig:qr}. Meanwhile, the green dashed line in Fig.~\ref{fig:qr} corresponds to the right-hand side of the equation. Evidently, the figure demonstrates that as the quench velocity approaches zero, the Berry curvature can be accurately approximated by the ratio of the response signal to the quench velocity.
To implement quantum sensing based on the dynamic response, we note that the quantity $\langle\psi(t_f)|S_y|\psi(t_f)\rangle$ can be measured in the experiment. By solving this non-linear equation, we can deduce the value of $D$ or, equivalently, the magnitude of the magnetic field $h$.
It is important to note that the quench process cannot cross the degenerate point.

\subsection{robustness to decoherence of the sensing protocol}
In this section, we examine the retrieval of the Berry curvature using quantum response in the presence of decoherence.
Building upon the numerical simulation presented in Sec.~\ref{sec:basic_scheme}, we extend our analysis to incorporate the influence of the environment, specifically the interaction with $N$ nuclear spins. This interaction is captured by the inclusion of the last term in Eq.~(\ref{eq:main_Hamiltonian}), which accounts for the coupling between the NV electronic spin and the nuclear spins.
The presence of coupling to the nuclear spins introduces decoherence effects on the NV electronic spin, particularly when the nuclear spins are partially polarized.
In this context, we consider the quenching process described by Eq.~(\ref{eq:rotating_quench_field}), and our objective is to calculate the response signal $M_y=\text{Tr}[\rho(t_f) S_y]$ at $t_f=\pi/v$, where $\rho(t_f)$ is the state of the compound system, $\rho(t_f)=U(t_f)\rho(0)U^\dagger(t_f)$.
The time evolution operator $U(t_f)=\mathcal{T}e^{-i\int_0^{t_f}H(t^\prime)dt^\prime}$, with $H(t)$ given by the Hamiltonian in Eq.~(\ref{eq:main_Hamiltonian}).
The initial state, $\rho(0)=|\phi_0\rangle\langle\phi_0|\otimes\rho_n$, consists of two components: $|\phi_0\rangle\langle\phi_0|$, which corresponds to the ground state of the NV Hamiltonian in the absence of coupling to the nuclear spin bath, and $\rho_n$, which represents the initial state of the nuclear spin bath. The nuclear spin bath is assumed to be in a thermal state and is characterized by the density matrix $\rho_{n}=(1/Z)\exp(-\beta \sum_{k=1}^NI_{kz})$. Here, $Z=[2\cosh(\beta/2)]^{N}$ represents the partition function, and $\beta=2\tanh^{-1}(P)$ denotes the inverse temperature, determined by the average nuclear polarization $P$~\cite{ding2014high}.

When dealing with a large number of nuclear spins ($N$), simulating the dynamics governed by a time-dependent Hamiltonian using the density matrix formalism becomes computationally challenging due to the exponential growth of the Hilbert space dimension ($\sim 2^{N+1}\times 2^{N+1}$).
To simplify the simulation for larger $N$, we employ certain approximations.
First, we assume a homogeneous coupling between the NV electronic spin and the nuclear spins, namely $\mathbf{A}_k=A$, based on the quasistatic approximation~\cite{zhang2006hyperfine,he2019exact}.
This allows us to utilize the collective nuclear spin operator $\mathbf{I}=\sum_{k=1}^N{\mathbf{I}_{k}}$, and the total angular momentum $\mathbf{J}=\mathbf{S}+\mathbf{I}$ becomes a constant of motion, leading to the reduction of the dimension of the Hilbert space.
Second, since the initial state of the nuclear spins is assumed to be in a thermal state, we can employ wave function dynamics instead of density matrix calculations.
Namely,
\begin{equation}
\rho(t_f)=\sum_{I_0=k}^{N/2}\sum_{M_0=-I_0}^{I_0}\omega(I_0,M_0)|\psi(t_f)\rangle\langle\psi(t_f)|,
\end{equation}
where $k=1/2$ if $N$ is odd, and $k=0$ if $N$ is even.
The time evolution of the wave function is given by $|\psi(t_f)\rangle=U(t_f)(|\phi_0\rangle\otimes|I_0,M_0\rangle)$, and the statistical weight associated with the nuclear spin state $|I_0,M_0\rangle$ is given by
\begin{equation}
\begin{aligned}
\omega&(I_0,M_0)\\
&=C_N^{\frac{N}{2}-I_0}(\frac{1+P}{2})^{\frac{N}{2}-M_0}(\frac{1-P}{2})^{\frac{N}{2}+M_0}\frac{2I_0+1}{\frac{N}{2}+I_0+1},
\end{aligned}
\end{equation}
where $C_N^M$ represents the binomial coefficient.
By employing these simplifications, we can tackle the simulation of the dynamics in a more computationally feasible manner while capturing the essential features of the system's behavior.

After making the simplifications mentioned earlier, we perform simulations of the quantum response experiment considering $N=20$ nuclear spins (with spin $I=1/2$).
The retrieved Berry curvatures for different average nuclear polarizations $P$ are presented in Fig.~\ref{fig:qr}.
The results reveal an interesting phenomenon: when the nuclear polarization is non-zero ($P=0.2$, red solid line), there exists an optimal quenching speed $v$ for extracting the Berry curvature using the quantum response formula. 
Several factors contribute to this observation. Firstly, according to adiabatic perturbation theory, a smaller quenching speed $v$ leads to a more accurate retrieval of the Berry curvature through quantum response. Secondly, a slower quenching speed implies a longer evolution time, increasing the impact of decoherence.
This competing mechanism leads to the existence of the optimal quench velocity.
However, a counterintuitive finding arises when the nuclear spins are completely unpolarized ($P=0$, blue solid line). Our calculations demonstrate that as the nuclear polarization approaches zero, indicating increased decoherence, the influence of decoherence on the quantum response experiment becomes less significant instead. 
This novel feature is in contrast to conventional Ramsey-based sensing schemes, where higher nuclear polarization is typically required to mitigate electronic spin decoherence.
The origin of this unique characteristic can be attributed to adiabatic perturbation theory, which suggests that the presence of decoherence or dephasing in the quantum system can actually enhance the applicability of the quantum response formula~\cite{Gritsev6457}.
This finding highlights the robustness of our quantum response-based sensing scheme to decoherence, making it highly feasible for realistic experiments, since the polarization of nuclear spins in solid-state systems is usually difficult and time-consuming~\cite{slichter2013principles}.

\section{vector magnetometry and motion sensing of magnetic nanoparticles}
\begin{figure}
\includegraphics[width=0.5\textwidth]{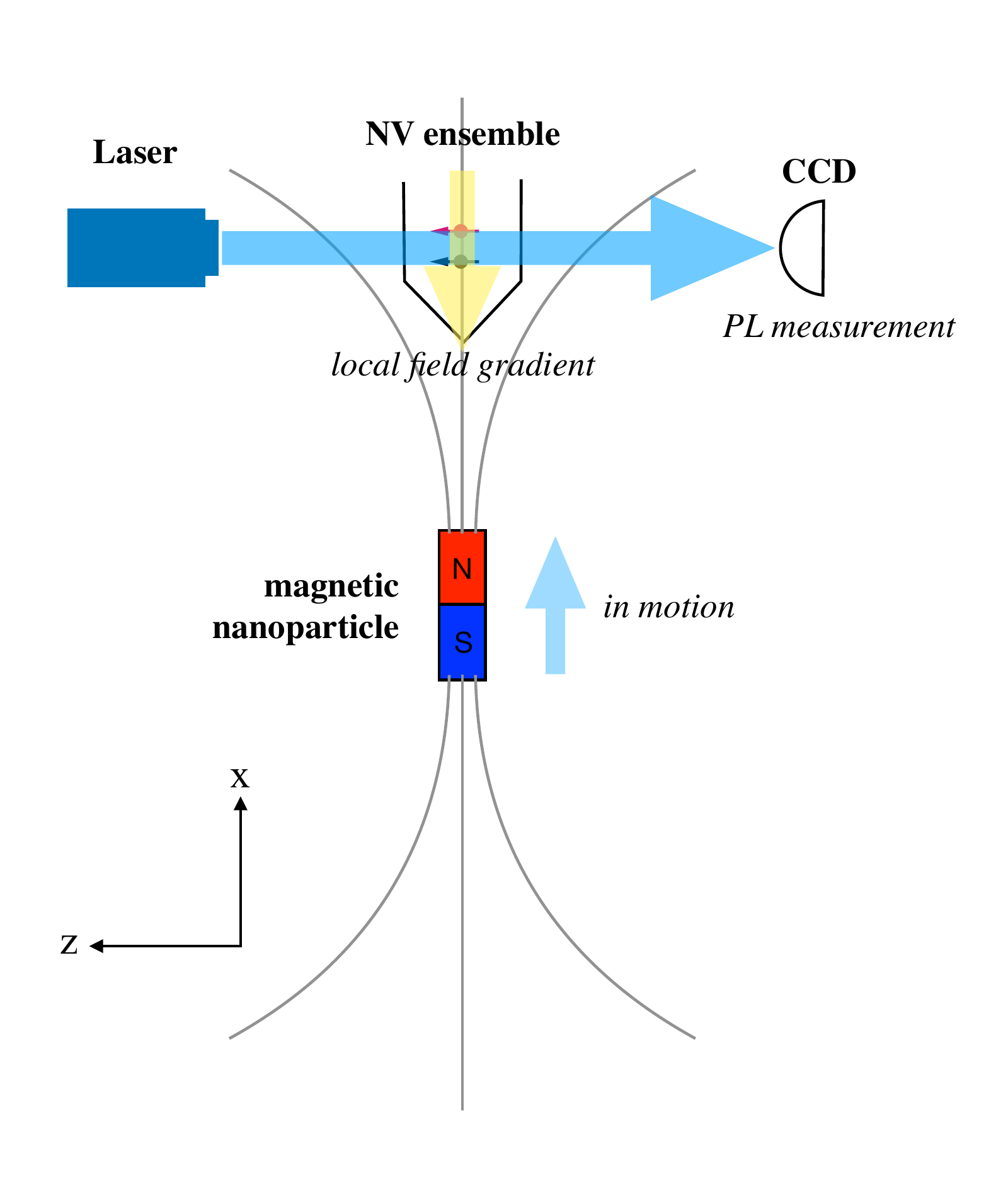}
\centering
\caption{\label{fig:nano_motion_sensing}Sensing the motion of the magnetic nanoparticle via quantum dynamic response. The NV ensemble in the diamond is optically initialized by illuminating it with a 532 nm laser, and the electronic spin state of the NV center can be determined by the spin state-dependent photoluminescence (PL). The static local field gradient can be introduced by mounting a nano magnet onto the diamond. Generally, the magnetic nanoparticle can be in the Brownian motion, but here we restrict the motion of the magnetic nanoparticle along the x-axis for clarity. }
\end{figure}

In recent years, several proposals have been put forward to realize vector magnetometry using solid-state spins~\cite{liu2019nanoscale, schloss2018simultaneous, lee2015vector, niethammer2016vector, zheng2020microwave}. In this section, we present a concrete example to demonstrate the implementation of vector magnetometry and the motion sensing of magnetic nanoparticles using NV centers in diamond through quantum dynamic response.
The schematic diagram in Fig.~\ref{fig:nano_motion_sensing} illustrates the setup, where NV centers are utilized to sense the motion of a magnetic nanoparticle and determine the instantaneous magnetic field generated by the magnetic nanoparticle itself~\cite{wang2018magnetic}. 
Typically, the magnetic nanoparticle undergoes Brownian motion, leading to a time-varying magnetic field experienced by the NV center. 
By formulating equations based on the quantum response formula, we can, in principle, determine the motion of the magnetic nanoparticle for arbitrary time dependencies, as long as the motion is nearly adiabatic.
Here, for clarity, we restrict the motion of the magnetic nanoparticle along the x-axis, to demonstrate the capability of the vector magnetometry and the motion sensing.
We now consider two ensembles of NV centers, and the Hamiltonian for the NV center in the $i$-th ensemble is given by:
\begin{equation}
H^{(i)}(t)=DS_z^2+h_z^{(i)}S_z+h_yS_y+h_x(t)S_x.
\end{equation}
Here, $h_z^{(i)}$ represents the static magnetic field applied to the $i$-th ensemble along the z-axis. 
Since these two ensembles of NV centers are usually close to each other, this different static field can be generated by mounting a nano magnet on the diamond.
The static magnetic field $h_y$ is common to both NV ensembles.
The static fields $h_z^{(i)}$ and $h_y$ are assumed to be known beforehand, which can be determined, for example, through conventional Ramsey-based magnetometry. 
The magnetic field $h_x(t)$, which we aim to detect, is generated by the magnetic nanoparticle. 
Initially, at time $t=0$, the magnetic nanoparticle is far away from the NV center, resulting in a negligible value for $h_x(t=0)$.
The initial state of the NV center is prepared in its ground state, which can be optically polarized by illuminating a 532 nm laser~\cite{hanson2008coherent,childress2006coherent}.

When the magnetic nanoparticle moves in close proximity to the NV center, the NV center experiences a time-varying magnetic field $h_x(t)$ along the x-axis. At a specific measurement time $t_f$, we perform measurements on the spin expectation values of the two NV ensembles, denoted as $\langle S_z^{(1)}\rangle$ and $\langle S_z^{(2)}\rangle$, utilizing spin state-dependent photoluminescence (PL)~\cite{hanson2008coherent}.
According to the quantum response formula, the relationship between these measured spin expectation values and the magnetic field components can be described by the following equations:
\begin{equation}
\begin{aligned}
\frac{\langle S_z^{(1)}\rangle}{v_x}=\mathcal{F}^{(1)}_{xz}[h_x(t_f),h_y,h_z^{(1)}],\\
\frac{\langle S_z^{(2)}\rangle}{v_x}=\mathcal{F}^{(1)}_{xz}[h_x(t_f),h_y,h_z^{(2)}],
\end{aligned}
\end{equation}
where the Berry curvature $\mathcal{F}_{xz}^{(1)}[h_x,h_y,h_z^{(i)}]$ is determined by Eq.~(\ref{eq:berry_curvature_cartesian}) and $v_x$ is the quench velocity.
By solving these nonlinear equations, we can obtain the instantaneous values of the magnetic field $h_x$ and the velocity $v_x$ at time $t_f$.
From the perspective of motion sensing, the proposed method allows us to determine the instantaneous velocity of the magnetic nanoparticle~\cite{cohen2020utilising}, and extract valuable information about its position by determining the magnetic field $h_x$.
The vector magnetometry can be realized in the same manner.
For instance, in the case where the value of $h_y$ is not known in advance, we can extend the setup by incorporating an additional ensemble of NV centers with a different $h_z^{(i)}$. This allows us to construct an additional nonlinear equation, enabling the determination of $h_y$ as well.
In other words, by constructing groups of these nonlinear equations using static field gradients, we eliminate the need to know the quench velocity beforehand to estimate the magnetic field.

In conclusion, we present a novel sensing proposal for detecting the motion of magnetic nanoparticles based on the mechanism of quantum dynamic response. 
This approach enables us to realize highly sensitive motion sensing within nanoscale, where the position and instantaneous velocity of the magnetic nanoparticle can be determined through the analysis of the measured spin expectation values. It offers a promising avenue for accurately tracking and characterizing the motion of nano-scale objects using solid-state spins.
This has significant implications in various fields, including bioimaging, where magnetic nanoparticles can serve as indicators for targeted imaging~\cite{balasubramanian2014nitrogen,wang2018magnetic}.

\section{discussion on the sensitivity}
In this section, we investigate the sensitivity of our dynamic response-based sensing scheme, specifically focusing on the sensing scheme discussed in Sec.~\ref{sec:scalar_rotating}. 
By analyzing the closed exact form of the Berry curvature given in Eq.~(\ref{eq:berry_curvature_spherical}), we can calculate the susceptibility of the Berry curvature with respect to the parameter $D$.
Remarkably, we can analytically calculate the susceptibility as $\theta$ approaches zero when $D=1$.
In this limit, the susceptibility exhibits the following behavior:
\begin{equation}
\begin{aligned}
\lim_{\theta\rightarrow 0}\frac{\partial}{\partial D}\left[\frac{8\sin^3\theta\cos^2\theta(D_1+D_2)^2(1-D_1D_2)}{\mathcal{N}_1\mathcal{N}_2(D_2-D_1)}\right]&=\infty,\\
\lim_{\theta\rightarrow 0}\frac{\partial}{\partial D}\left[\frac{8\sin^3\theta\cos^2\theta(D_1+D_3)^2(1-D_1D_3)}{\mathcal{N}_1\mathcal{N}_3(D_3-D_1)}\right]&=-\frac{1}{8\sqrt{2}},
\end{aligned}
\end{equation}
which indicates that,
\begin{equation}
\label{eq:divergent_susceptibility}
\lim_{\theta\rightarrow 0}\frac{\partial \mathcal{F}_{\phi\theta}^{(1)}}{\partial D}=\infty.
\end{equation}
Apparently, this indicates that near the work point $(\theta=0,D=1)$, a slight change in $D$ will result in a significant variation in the Berry curvature $\mathcal{F}_{\phi\theta}^{(1)}$, which corresponds to a measurable quantity divided by the quench velocity in the experiment.
Consequently, we anticipate an exceptionally high sensitivity near the work point in our dynamic response sensing scheme.
This is reminiscent of the sensor utility of non-Hermitian systems, where the susceptibility of certain measurable quantities can also exhibit divergent behaviors~\cite{chu2020quantum}.

However, it is important to note that the work point $(\theta=0, D=1)$ actually corresponds to an energy degenerate point ($E_1=E_2$).
Thus, achieving near adiabatic conditions when approaching this point requires an extremely small quenching velocity. 
Consequently, while the susceptibility near the work point may be divergent, it is accompanied by a significantly longer evolution time. 
Therefore, the divergence in susceptibility does not necessarily translate into a divergence in sensitivity.
In fact, a general bound for the estimation uncertainty has been proposed in Ref.~\cite{rams2018at} for dynamic quantum sensing schemes, taking into account the evolution time explicitly. 
When the parameter encoding process (for both sudden quench and adiabatic quench) is governed by the parameter Hamiltonian $\hat{H}_\lambda$, this bound is given by
\begin{equation}
\label{eq:ultimate_sensitivity_bound}
\delta\lambda\geq\frac{1}{t||\frac{\partial \hat{H}_\lambda}{\partial\lambda}||},
\end{equation}
where $||\hat{A}||$ represents the seminorm defined as the difference between the maximum and minimum eigenvalues of the operator $\hat{A}$, i.e., $||\hat{A}||=E_{\text{max}}-E_{\text{min}}$.
In the dynamic sensing protocol described in Sec. III, the ultimate sensitivity bound for estimating the parameter $D$ is given by $\delta D\geq 1/t$.

Since both our dynamic response-based sensing scheme and the conventional Ramsey-based sensing scheme are subject to the same ultimate sensitivity bound as described by Eq.~(\ref{eq:ultimate_sensitivity_bound}), the divergence in the susceptibility of the Berry curvature presented in Eq.~(\ref{eq:divergent_susceptibility}) does not necessarily imply a divergent sensitivity. Hence, our dynamic response-based sensing scheme does not offer an inherently enhanced ultimate sensitivity compared to the Ramsey-based scheme.
However, the advantage of our dynamic response-based sensing scheme lies in its capability to sense time-varying magnetic fields or the motion of magnetic nanoparticles, which remains challenging for conventional interference-based sensing schemes. 
This opens up new possibilities for applications in dynamic sensing scenarios where conventional schemes fall short.

\section{Summary}
The essence of our dynamic response-based sensing scheme lies in utilizing the dynamics governed by a \textit{time-dependent} Hamiltonian to encode the parameter of interest into the quantum state.
Usually, calculating the dynamics governed by a time-dependent Hamiltonian, like using the time-ordering evolution operator, can be challenging, limiting its application in quantum sensing.
However, the quantum response theory offers a valuable tool by providing a simple and clear expression of the observable dynamics in terms of the Berry curvature, as long as the time dependence of the Hamiltonian is near adiabatic.
In this study, we leverage this relation to demonstrate the power of the quench-response mechanism in realizing quantum sensing. 
Unlike conventional interference or Ramsey-based sensing schemes, which rely on time-independent Hamiltonians to encode the parameter, our dynamic response-based sensing scheme offers distinct advantages.
It enables the sensing of instantaneous magnetic fields and the detection of the motion of magnetic nanoparticles.
This capability opens up new possibilities in quantum sensing, particularly in scenarios where the parameter to be estimated are time-dependent and require real-time measurements.

In this study, we employ the NV center in diamond as our platform to demonstrate the effectiveness of the dynamic response-based sensing scheme. 
By analytically deriving the exact form of the Berry curvature, we are able to design quench-response protocols that enable us to accurately estimate the magnitude of the magnetic field or the quench velocity.
One of the notable advantages of our dynamic response-based sensing scheme is its robustness to decoherence. 
Contrary to conventional interference-based approaches, we find that a vanishing nuclear polarization actually benefits our scheme.
This counterintuitive result highlights the unique properties of the dynamic response-based approach and its resilience to decoherence effects. This robustness is a significant advantage, making our scheme highly feasible for realistic experiments.
Furthermore, by exploiting the quench-response mechanism, we propose schemes that enable the detection and characterization of the motion of magnetic nanoparticles. This advancement opens up new possibilities for applications in bioimaging and other areas where accurate motion tracking within nanoscale is essential.

In fact, the principle of our dynamic sensing scheme can be extended to other quantum systems, including quantum many-body systems, whether they are interacting or not. 
While the exact form of the Berry curvature may not be obtainable in these systems, it can still be measured experimentally through alternative methods~\cite{bleu2018effective,ozawa2018extracting} or via the quantum response theory introduced here.
By measuring the value of the Berry curvature in advance, we can design dynamic sensing protocols to detect the quench velocity in these systems.
The dynamic response-based sensing scheme proposed in this study offers the advantage of technical simplicity, making it highly accessible for practical implementation in experimental settings.
Our study demonstrates the potential of utilizing the dynamic response and the quench-response mechanism to realize a novel sensing scheme. However, there are still untapped possibilities and further potentials to explore in the field of quantum sensing using this approach. Future research can delve deeper into these unexplored avenues and uncover new applications and insights.

\begin{acknowledgments}
This work was supported by the National Key Research and Development Program of China (Grants No. 2017YFA0304202 and No. 2017YFA0205700), the NSFC through Grant No. 11875231 and No. 11935012, and the Fundamental Research Funds for the Central Universities through Grant No. 2018FZA3005.
\end{acknowledgments}

\appendix
\section{Review of the quantum response theory}
To render this work more self-consistent, we now make a brief review on the adiabatic perturbation theory and the quantum response theory.
More details can be found in Refs.~\cite{Gritsev6457,de2010adiabatic,rigolin2008beyond}.
The Schr\"{o}dinger equation for a time-dependent Hamiltonian is
\begin{equation}
i\frac{\partial |\psi(t)\rangle}{\partial t}=H(t)|\psi(t)\rangle.
\end{equation}
Here we expand the wave function using the instantaneous eigenstates as
\begin{equation}
|\psi(t)\rangle=\sum_n a_n(t)|\phi_n(t)\rangle,
\end{equation}
with $H(t)|\phi_n(t)\rangle=E_n(t)|\phi_n(t)\rangle$.
Thus, the Schr\"{o}dinger equation can be represented as (by left multiplying $\langle \phi_m(t)|$ on both sides),
\begin{equation}
i\frac{\partial a_m(t)}{\partial t}+i\sum_n a_n(t)\langle \phi_m(t)|\frac{\partial }{\partial t}|\phi_n(t)\rangle=E_m(t)a_m(t).
\end{equation}
We now make the gauge transformation $a_n(t)=\alpha_n(t)e^{-i\omega_n(t)}e^{i\gamma_n(t)}$, where the dynamic phase is defined as $\omega_n(t)\equiv-\int_{t}^{t_f}{E_n(\tau)d\tau}$, and the Berry phase is defined as $\gamma_n(t)=-i\int_{t}^{t_f}{\langle n|\frac{\partial}{\partial t^\prime}|n\rangle dt^\prime}$.
As a result, we obtain (the indices $m\leftrightarrow n$ are exchanged)
\begin{equation}
\frac{\partial \alpha_n(t)}{\partial t}=-\sum_{m\neq n}{\alpha_m(t)\langle \phi_n(t)|\frac{\partial}{\partial t}|\phi_m(t)\rangle e^{i(\omega_{nm}(t)-\gamma_{nm}(t))}}, 
\end{equation}
where $\omega_{nm}(t)=\omega_n(t)-\omega_m(t)$ and $\gamma_{nm}(t)=\gamma_n(t)-\gamma_m(t)$.
Alternatively, we can write it in the integral form as follows:
\begin{equation}
\begin{aligned}
&\alpha_n(t)=\\
&-\int_{t_i}^t{dt^\prime}\sum_{m\neq n}{\alpha_m(t^\prime)\langle \phi_n(t^\prime)|\frac{\partial}{\partial t^\prime}|\phi_m(t^\prime)\rangle e^{i(\omega_{nm}(t^\prime)-\gamma_{nm}(t^\prime))}}.
\end{aligned}
\end{equation}
Now if the initial state is in the ground state, namely, $\alpha_0(0)=1$ and $\alpha_m(0)= 0$ for $m \neq 0$, by making the adiabatic perturbation approximation~\cite{de2010adiabatic,rigolin2008beyond}, we obtain
\begin{equation}
\alpha_n(t)\approx -\int_{t_i}^t{dt^\prime}{\langle \phi_n(t^\prime)|\frac{\partial}{\partial t^\prime}|\phi_0(t^\prime)\rangle e^{i(\omega_{n0}(t^\prime)-\gamma_{n0}(t^\prime))}}.
\end{equation} 
Using integration by parts, we obtain that,
\begin{equation}
\begin{aligned}
&\alpha_n{(t_f)}\approx\left[ i\frac{\langle \phi_n(t)|\frac{\partial}{\partial t}|\phi_0(t)\rangle}{E_n(t)-E_0(t)}\right.\\
&\left.-\frac{1}{E_n(t)-E_0(t)}\frac{d}{dt}\frac{\langle \phi_n(t)|\frac{\partial}{\partial t}|\phi_0(t)\rangle}{E_n(t)-E_0(t)}+\ldots\right]\times\\
&\left.e^{i(\omega_{n0}(t)-\gamma_{n0}(t))}\right|_{t_i}^{t_f}.
\end{aligned}
\end{equation}
Since the time dependence of the Hamiltonian is usually introduced through the time varying parameter, namely, $H(t)\equiv H(\lambda(t))$, we have the following relation,
\begin{equation}
\langle \phi_n(t)|\frac{\partial}{\partial t}|\phi_0(t)\rangle=\frac{\partial\lambda}{\partial t}\langle \phi_n(\lambda)|\frac{\partial}{\partial \lambda}|\phi_0(\lambda)\rangle,
\end{equation}
while $\omega_n(\lambda)\equiv-\int_{\lambda}^{\lambda_f}{\frac{E_n(\lambda^\prime)}{v(\lambda^\prime)}d\lambda^\prime}$, with $v(\lambda)=\frac{d\lambda}{dt}$, and $\gamma_n(\lambda)=-i\int_{\lambda}^{\lambda_f}{\langle n|\frac{\partial}{\partial \lambda^\prime}|n\rangle d\lambda^\prime}$.
Therefore, the integral above can be rewritten as follows:
\begin{equation}
\begin{aligned}
\alpha_n&{(\lambda_f)}\approx\left[ i\frac{\partial \lambda}{\partial t}\frac{\langle \phi_n(\lambda)|\frac{\partial}{\partial \lambda}|\phi_0(\lambda)\rangle}{E_n(\lambda)-E_0(\lambda)}-\frac{\partial^2 \lambda}{\partial t^2}\frac{\langle \phi_n(\lambda)|\frac{\partial}{\partial \lambda}|\phi_0(\lambda)\rangle}{[E_n(\lambda)-E_0(\lambda)]^2}\right.\\
&\left.-(\frac{\partial \lambda}{\partial t})^2\frac{1}{E_n(\lambda)-E_0(\lambda)}\frac{d}{d\lambda}\frac{\langle \phi_n(\lambda)|\frac{\partial}{\partial \lambda}|\phi_0(\lambda)\rangle}{E_n(\lambda)-E_0(\lambda)}+\ldots\right]\times\\
&\left.e^{i(\omega_{n0}(\lambda)-\gamma_{n0}(\lambda))}\right|_{\lambda_i}^{\lambda_f}.
\end{aligned}
\end{equation}
When the quench is near adiabatic ($\frac{\partial \lambda}{\partial t}\rightarrow 0$), the transition amplitude can be approximated as
\begin{equation}
\begin{aligned}
\alpha_n&{(\lambda_f)}\approx i\left.\frac{\partial \lambda}{\partial t}\frac{\langle \phi_n(\lambda)|\frac{\partial}{\partial \lambda}|\phi_0(\lambda)\rangle}{E_n(\lambda)-E_0(\lambda)}e^{i(\omega_{n0}(\lambda)-\gamma_{n0}(\lambda))}\right|_{\lambda_i}^{\lambda_f}.
\end{aligned}
\end{equation}
Particularly, when the energy gap is large or the quench velocity is vanishing at the initial time, we have
\begin{equation}
\begin{aligned}
a_n{(\lambda_f)}&=\alpha_n{(\lambda_f)}e^{-i\omega_n(\lambda_f)}e^{i\gamma_n(\lambda_f)}\\
&\approx i\left.\frac{\partial \lambda}{\partial t}\frac{\langle \phi_n(\lambda)|\frac{\partial}{\partial \lambda}|\phi_0(\lambda)\rangle}{E_n(\lambda)-E_0(\lambda)}\right|_{\lambda_f}.
\end{aligned}
\end{equation}
This is the result presented in Ref.~\citep{Gritsev6457}.
We can also utilize the following relation:
\begin{equation}
\langle \phi_n(\lambda)|\frac{\partial}{\partial\lambda}|\phi_m(\lambda)\rangle=-\frac{\langle\phi_n(\lambda)|\frac{\partial H}{\partial \lambda}|\phi_m(\lambda)\rangle}{E_n(\lambda)-E_m(\lambda)}.
\end{equation}
Thus, we have the response signal along the $\mu$-direction as a function of the quench velocity $v_\lambda\equiv \frac{\partial \lambda}{\partial t} $ up to the leading order as follows:
\begin{equation}
\begin{aligned}
M_\mu&\equiv-\langle \psi(t_f)|\frac{\partial H}{\partial \mu}|\psi(t_f)\rangle\approx -\langle \phi_0|\frac{\partial H}{\partial \mu}|\phi_0\rangle\\
&+\left.i\frac{\partial \lambda}{\partial t}\sum_{n\neq 0}\frac{\langle \phi_0|\frac{\partial H}{\partial \mu}|\phi_n\rangle\langle \phi_n|\frac{\partial H}{\partial \lambda}|\phi_0\rangle-\mu\leftrightarrow\lambda}{[E_n(\lambda)-E_0(\lambda)]^2}\right|_{\lambda_f}.
\end{aligned}
\end{equation}
This leads to the general formula of the quantum response as follows:
\begin{equation}
\label{eq:qr_formula}
M_\mu=\text{const}+v_\lambda\mathcal{F}^{(0)}_{\mu\lambda}+\mathcal{O}(v_\lambda^2),
\end{equation}
where the Berry curvature is given by
\begin{equation}
\label{eq:berry_curvature_general_a}
\mathcal{F}_{\mu\lambda}^{(m)}=i\sum_{n\neq m}\frac{\langle \phi_m|\frac{\partial H}{\partial \mu}|\phi_n\rangle\langle \phi_n|\frac{\partial H}{\partial \lambda}|\phi_m\rangle-\mu\leftrightarrow\lambda}{[E_n(\lambda)-E_m(\lambda)]^2}.
\end{equation}

\section{Exact eigenvalues and eigenvectors of a $3\times 3$ Hermitian matrix}
In this section, we provide the analytic solution of the eigenvalues and eigenvectors of a general $3\times 3$ Hermitian matrix represented as follows:
\begin{equation}
H=\begin{pmatrix}
a_{11} & a_{12} & a_{13} \\
a_{12}^* & a_{22} & a_{23} \\
a_{13}^* & a_{23}^* & a_{33}
\end{pmatrix}.
\end{equation}
The secular equation to calculate the eigenvalue is
\begin{equation}
\det{(H-\lambda\mathbf{1})}=0,
\end{equation}
which, according to the Cayley-Hamilton theorem, corresponds to the cubic equation
\begin{equation}
\lambda^3-\text{Tr}(H)\lambda^2-\frac{1}{2}[\text{Tr}(H^2)-(\text{Tr}(H))^2]\lambda-\det(H)=0.
\end{equation}
Since $H$ is a Hermitian operator, $\text{Tr}(H^2)$, $\text{Tr}(H)$ and $\det(H)$ are all real quantities.
To further simplify the corresponding cubic equation, we now make some transformations as follows:
\begin{equation}
\begin{aligned}
B&=H-\frac{\text{Tr}[H]}{3}\mathbf{1},\\
A&=\sqrt{\frac{2}{\text{Tr}[B^2]}}B.
\end{aligned}
\end{equation}
As a result, the eigenvalues of $H$ and the eigenvalues of $A$ follow the relation:
\begin{equation}
\label{eq:relation_eigenvalue}
\lambda_k=\sqrt{\frac{\text{Tr}[B^2]}{2}}t_k+\frac{\text{Tr}[H]}{3}.
\end{equation}
We notice that $\text{Tr}[A]=0$ and $\text{Tr}[A^2]=2$.
Consequently, the secular equation to calculate the eigenvalues of $A$ becomes a depressed cubic equation:
\begin{equation}
\label{eq:cubic_eq1}
t^3-t-q=0,
\end{equation}
with $q=\det(A)$.
Since the operator $A$ is still a Hermitian operator, all the eigenvalues are real, then we can assume the solution to be $t=u\cos\theta$.
We can prove that $-\frac{2}{3\sqrt{3}}<q<\frac{2}{3\sqrt{3}}$, when Eq.~(\ref{eq:cubic_eq1}) has three distinct real roots (it is easy to observe by plotting the graph of the function).
Specifically, when $q=\frac{2}{3\sqrt{3}}$, two multiple roots correspond to the stationary point of $f(t)=t^3-t$, namely $t_1=t_2=\frac{1}{\sqrt{3}}$, and $t_3=-\frac{2}{\sqrt{3}}$.
It is similar when $q=-\frac{2}{3\sqrt{3}}$, and we can conclude that $-\frac{2}{\sqrt{3}}\leq t\leq \frac{2}{\sqrt{3}}$.
As a result, we can choose $u=\frac{2}{\sqrt{3}}$.
After dividing the equation by $u^3/4$, the depressed cubic equation in Eq.~(\ref{eq:cubic_eq1}) now becomes,
\begin{equation}
4\cos^3\theta-3\cos\theta-\frac{3}{2}\sqrt{3}q=0.
\end{equation}
Using the trigonometric identity
\begin{equation}
4\cos^3\theta-3\cos\theta=\cos(3\theta),
\end{equation}
we obtain that,
\begin{equation}
\cos(3\theta)=\frac{3}{2}\sqrt{3}q.
\end{equation}
As a result, we have the three eigenvalues of matrix $A$ as follows:
\begin{equation}
t_k=\frac{2}{\sqrt{3}}\cos[\frac{1}{3}\arccos(\frac{3}{2}\sqrt{3}\det(A))-\frac{2\pi k}{3}],
\end{equation}
for $k=0,1,2$.
Then, the eigenvalues of $H$ can be determined by Eq.~(\ref{eq:relation_eigenvalue}).

The eigenstates of the $3\times 3$ Hermitian matrix $H$ can be represented as the cross product of two three-dimensional vectors, $|\tilde{\Psi}_m\rangle=[(\mathbf{h}_1-E_m\mathbf{e}_1)\times(\mathbf{h}_3-E_m\mathbf{e}_3)]^*$, as long as the two vectors are linear independent~\cite{kopp2008efficient}.
Here, $\mathbf{h}_j$ is the $j-$th column of the Hermitian matrix $H$, and $\mathbf{e}_i$ is the unit vector, like $\mathbf{e}_1=(1,0,0)^{\text{T}}$.
We now make a brief proof to show that $|\tilde{\Psi}_m\rangle$ is indeed the eigenstate.
First, if $|\tilde{\Psi}_m\rangle$ is the eigenstate, then we have $(H-E_m\mathbf{1})|\tilde{\Psi}_m\rangle=0$, or equivalently we have to prove that,
\begin{equation}
\langle \tilde{\Psi}_m|(H-E_m\mathbf{1})|\psi\rangle=0,
\end{equation}
where $|\psi\rangle=\alpha_1 \mathbf{e}_1+\alpha_2 \mathbf{e}_2+\alpha_3 \mathbf{e}_3$ is an arbitrary wave vector.
After the expansion, we have
\begin{equation}
\begin{aligned}
\langle \tilde{\Psi}_m &|(H-E_m\mathbf{1})|\psi\rangle \\
=&\alpha_1\langle \tilde{\Psi}_m|(\mathbf{h}_1-E_m\mathbf{e}_1)\rangle+\alpha_2\langle \tilde{\Psi}_m|(\mathbf{h}_2-E_m\mathbf{e}_2)\rangle\\
&+\alpha_3\langle \tilde{\Psi}_m|(\mathbf{h}_3-E_m\mathbf{e}_3)\rangle.
\end{aligned}
\end{equation}
Obviously, both the first term and the last term equal zero. To prove the second term equals zero, we need to utilize the property of the mixed product as follows:
\begin{equation}
(\mathbf{a}\times\mathbf{b})\cdot\mathbf{c}=\det\begin{pmatrix}
a_{1} & b_{1} & c_{1} \\
a_{2} & b_{2} & c_{2} \\
a_{3} & b_{3} & c_{3}
\end{pmatrix}.
\end{equation}
As a result, we can prove that
\begin{equation*}
\begin{aligned}
\langle \tilde{\Psi}_m&|(\mathbf{h}_2-E_m\mathbf{e}_2)\rangle\\
&=[(\mathbf{h}_1-E_m\mathbf{e}_1)\times(\mathbf{h}_3-E_m\mathbf{e}_3)]\cdot(\mathbf{h}_2-E_m\mathbf{e}_2)\\
&=\det(H-E_m\mathbf{1})=0.
\end{aligned}
\end{equation*}

When these two vectors are linear dependent, namely, $(\mathbf{h}_1-E_m\mathbf{e}_1)=\mu(\mathbf{h}_3-E_m\mathbf{e}_3)$, the eigenstate can be straightforwardly calculated by solving $(H-E_m\mathbf{1})|\tilde{\Psi}_m\rangle=0$ and the normalized eigenstate is given by
\begin{equation}
|\Psi_m\rangle=\frac{1}{1+|\mu|^2}\left(
\begin{array}{ccc}
1\\
0\\
-\mu
\end{array}\right).
\end{equation}
For instance, this is the situation when $\theta=\pi/2$ in the Hamiltonian [Eq.~(\ref{eq:simple_Hamiltonian})] in the main text, where for $E_2=D$, the corresponding eigenstate can be determined using the above expression.


%

\end{document}